\documentclass[jou]{apa7}
\usepackage{lipsum}
\usepackage[american]{babel}
\usepackage{csquotes}
\usepackage[style=apa,sortcites=true,sorting=nyt,backend=biber]{biblatex}
\DeclareLanguageMapping{american}{american-apa}
\usepackage{graphicx}
\usepackage{tikz}
\usepackage{amsmath}
\usepackage{amssymb}
\usepackage{hyperref}
\usetikzlibrary{arrows.meta, positioning}

\title{Measurement Induced Confounding}
\shorttitle{Measurement Induced Confounding}
\leftheader{Perrett and Kanopka}
\author{\addORCIDlink{George Perrett}{0000-0002-1930-2581} and \addORCIDlink{Klint Kanopka}{0000-0003-3196-9538}}
\authorsaffiliations{New York University}

\abstract{A critical assumption of observational studies is that all confounding variables must be known and sufficiently adjusted for to estimate causal effects. An implicit, and often overlooked, aspect of this assumption is that all confounding variables have been measured without error. In the social and medical sciences, latent traits such as motivation, self-efficacy, and ability measures are likely confounding variables. Because latent traits are not directly observable, conventional approaches to adjust for them in observational studies rely on collecting responses to individual items on a test or survey instrument and then adjust for sum scores, measurement model-derived ability estimates, or item responses directly. Through a process we describe as \textit{measurement induced confounding}, we show that measurement error propagates through the estimation process and that current conventional approaches to adjusting for latent traits in observational studies produce biased estimates of the average treatment effect with incorrectly calibrated coverage properties. A critical implication of this finding is that current observational studies that attempt to adjust for latent confounding variables likely put forth biased causal estimates with incorrect uncertainty intervals. We show that measurement induced confounding can be resolved through a Bayesian Joint Estimation approach that simultaneously estimates the measurement model, the treatment assignment model, and the response model.}


\authornote{Correspondence concerning this article should be addressed to George Perrett, Department of Applied Statistics, Social Science, and Humanities, New York University, 246 Greene St., Floor 3, New York, NY, 10003.  E-mail: \href{mailto:gp77@nyu.edu}{gp77@nyu.edu}}

\begin{document}
\maketitle

\section{Introduction}

When randomized studies are ethically or logistically infeasible, researchers have no choice but to turn to observational studies to answer causal questions. The primary limitation of observational studies is the assumption that all confounding variables are known and appropriately conditioned on. Work on propensity score methods \parencite{austin2011introduction, rosenbaum1984reducing, dehejia2002propensity} and recent advances in machine learning for causal inference \parencite{hill2011bayesian, hahn2020bayesian,hill2023machine} have focused on sufficiently conditioning on measured confounding variables without making unnecessary parametric modeling assumptions. An overlooked subtlety of these approaches is that they not only assume all confounding variables are measured \textit{but that they have been measured without error}. Methodological work in causal inference often begins from the position that there is zero measurement error in all confounding variables \parencite{dorie2019automated}. In practice, assuming all confounders are measured without error is problematic because confounding variables in the social, educational, and medical sciences commonly include measures of latent traits, including---but not limited to---mathematical ability \parencite{olitsky2014academic}, well-being \parencite{vandecandelaere2016effects}, personality \parencite{wagner2015first}, depression \parencite{thibodeau2013anxiety}, and motivation \parencite{kool2017effects}. Methodological work in psychometrics specifically concerned with the measurement of these sorts of latent constructs explicitly assumes that latent traits are measured with error. 

To illustrate the problem posed by confounding variables that are latent traits, we will consider a hypothetical observational study on the causal effect of attending college on income. Knowing the causal effect of college education on income is a pressing policy question with great societal implications, yet running a randomized study to answer it is impossible, leaving us no choice but to rely on observational methods.\footnote{Randomizing students to either attend or not attend college is ethically non-permissible.} For simplicity, suppose we knew that motivation was the only confounding variable and student motivation was linearly related to the probability that they attend college and also linearly related to their income. Here motivation meets the classical definition of a confounding variable and failing to sufficiently adjust for motivation would bias our estimate of the causal effect of college attendance on income. 

As such, in order to estimate the causal effect of college attendance on income, we must sufficiently adjust for motivation. The problem we face is that motivation is a latent trait that can not be directly observed. To measure motivation, we rely on instruments such as the achievement motivation scale \parencite{cassidy1989multifactorial} or Motivational Trait Questionnaire \parencite{heggestad2000individual}. In our hypothetical observational study, we will imagine we have collected responses to a 100-item instrument designed to measure motivation in this population. 


Given our 100 item responses per subject, there are multiple strategies for attempting to adjust for motivation. A reasonable first approach is computing a sum score by adding each individual's 100-item responses together. This method is commonly used in the social sciences but it does not consider the possibility that some items may reveal more about an individual's motivation than others. Item Response Theory \parencite[IRT;][]{lord2008statistical} provides a framework that allows discrete items with categorical responses to provide continuous estimates of latent motivation. Adjusting for an IRT estimate of motivation offers greater flexibility than the sum-score approach, as it also allows items to contribute differential amounts of information. While more flexible IRT models and sum scoring differ in how they leverage information from item responses, they are similar in that they both invoke a model to collapse information from many items into a single score of the latent trait of interest. A recent branch of psychometrics work has advocated for directly using item information and avoid collapsing information into a single metric \parencite{bruhn2025test, gilbert2025estimating}, and following these lines of research, rather than adjusting for a single summary of the item response string, we could also attempt to control for motivation by adjusting for the responses to all 100 items directly.

In this work, we argue that adjusting for sum scores, IRT-derived scores, or even individual item responses themselves are all insufficient to condition on confounding from latent traits. Through a process we describe as \textit{measurement induced confounding}, measurement error propagates through the adjustment process, resulting in biased estimates and undercoverage of causal effects. An uncomfortable implication of this work is that any observational study that uses conventional approaches to adjust for latent trait confounders may present both biased parameter estimates and has incorrect coverage properties. Fortunately, there is a path forward: We propose using Bayesian Joint Estimation of the measurement model, the response surface, and the assignment mechanism to address measurement-induced confounding. 

Before we can formalize the problem of measurement induced confounding and outline Bayesian Joint Estimation as a solution, we first begin by providing a brief overview of Item Response Theory and causal inference.

\section{Measurement and Item Response Theory}

A common challenge in behavioral research and the social sciences is the measurement of \textit{latent traits}, or qualities of an individual that are not themselves directly observable. Typically, the levels of these traits need to be inferred from something that is directly observable (e.g., responses to individual items on a test or survey, performances in structured tasks, etc.). We refer to any individual piece of observed information as an item and represent individual $i$'s response to item $j$ as $X_{ij}$, denoting an individual realization of that response as $x_{ij}$. In the case of dichotomous items that are scored as correct/incorrect (or endorsed/not endorsed), we have $X_{ij} \in \{0,1\}$, with $X_{ij}=1$ when individual $i$ responds correctly to item $j$. From this collection of responses, we aim to estimate individual $i$'s latent ability, $\theta_i$, from the information contained in the $J$-length vector of their item responses, $\mathbf{x_i} = \{x_{i1}, \ldots, x_{iJ}\}$. The model we use for this estimation is referred to as a \textit{measurement model}.

Let us first consider the sum score,
\begin{equation*}
    \hat{\theta}_i^{sum} = \sum_{j=1}^J x_{ij},
\end{equation*}
noting that it is itself a measurement model that carries strong assumptions, most obviously that each individual item contributes equally to the estimate of the latent trait. While this assumption may hold for carefully designed and tested instruments, it is frequently implausible. Additionally, when comparing sum scores across individuals, if test forms are not identical across all examinees, they must be the same length and assumed to be exchangeable; that is to say, a person's expected score should be independent of the specific form they were administered.

An alternative measurement model commonly employed in educational testing is IRT. Instead of performing bulk aggregation of individual response vectors, IRT posits that individual item responses are observed with some error and that the probability of observing a correct response arises from an interaction between the latent trait of the individual and some quality of the item itself. While a litany of IRT models exist, we choose to focus here on the \textit{two parameter logistic} model \parencite[2PL; for a broad survey of others, see][]{tutz2025short}. The 2PL gets its name from the fact that it models the probability of correct response conditional on an individual's latent ability using \textit{two parameters} to represent item features that are mapped onto a probability using a \textit{logistic} link function. That is,
\begin{equation}
    P(X_{ij} = 1\mid \theta_i) = \frac{1}{1+e^{-a_j(\theta_i - b_j)}},
\end{equation}
where $b_j$ is referred to as the item's \textit{difficulty} and $a_j$ is the item's \textit{discrimination}. A key feature of IRT is that items and respondents are placed onto the same scale, and the probability of correct response is a function of the signed distance between and individual's location on that scale, $\theta_i$, and an item's location on that scale, $b_j$, weighted by the item's discrimination $a_j$. In the case where $\theta_i=b_j$, $P(X_{ij}\mid\theta_i)=0.5$, linking interpretations between the placement of different objects on a common scale. This means that when $\theta_i > b_j$, $P(X_{ij}\mid\theta_i)>0.5$, so that individuals are more likely to answer items below their position on the scale correctly, and items above their position on the scale incorrectly. The discrimination controls the slope of the response curve at its midpoint; this is the rate at which the probability of correct response scales with changes in latent ability. Discrimination can be thought of as the amount of certainty you have in an item's ability to discriminate between high and low ability respondents. When $a_j=0$, $P(X_{ij}=1\mid\theta_i)=0.5$ for all values of $\theta_i$. In this way, the item provides no information about individual ability. Conversely, as $a_j \rightarrow \infty$, when $\theta_i > b_j$ the probability of correct response $P(X_{ij}=1\mid\theta_i) \rightarrow 1$ (and the probability of correct response approaches zero when $\theta_i < b_i)$. 

When using a 2PL IRT model as specified above, there are three core assumptions that are made: monotonicity, unidimensionality, and local independence. Monotonicity refers to the fact that probability of correct response is strictly a monotonic function of latent ability. Unidimensionality refers to the fact that probability of correct response relies on (or nearly relies on) a single latent trait. Finally local independence refers to the assumption that individual item responses are considered to be independent, conditional on latent ability.\footnote{Note that violations of these assumptions are not dealbreakers and can be accounted for with alternative model specifications; for examples, see \parencite{van1997item}} When these assumptions are met, IRT produces trait estimates that are comparable even if individuals respond to different items or different numbers of items, making it an attractive measurement model in the presence of missing data. Estimation is typically carried out by first estimating item parameters using marginal maximum likelihood estimation using an Expectation-Maximization procedure \parencite{bock1981marginal}, with estimation of individual latent abilities done after.

Moving forward, we use $\hat{\theta}^{sum}$ to represent estimates of the latent trait from a sum score measurement model and $\hat{\theta}^{IRT}$ to represent estimates of the latent trait from a 2PL IRT measurement model. 

\section{Causal Inference Notation, Estimands, and Assumptions}

\subsection{Notation and Causal Estimands}

Causal effects are the difference between potential outcomes \parencite{holland1986statistics}. Let $Z$ represent a binary treatment indicator. For a given individual $i$,  $Z_i = 1$ indicates that individual $i$ received the treatment, and $Z_i = 0$ indicates that individual $i$ did not receive the treatment. In the context of our college example, $Z_i = 1$ denotes someone who attended college, and $Z_i = 0$ denotes someone who did not attend college. Under a binary treatment, each individual has two potential outcomes: $Y_i(1)$ is what the outcome for individual $i$ would have been under the treatment, and $Y_i(0)$ is what the outcome for individual $i$ would have been without the treatment. In our context, $Y_i(1)$ represents what an individual's earnings would have been if they had gone to college, and $Y_i(0)$ represents what that same individual's earnings would have been if they did not attend college. 

An individual's causal effect is the difference between that individual's potential outcomes. We can formalize the individual causal effect as $\tau$. For a given individual $i$, their causal effect is $\tau_i = Y_i(1) - Y_i(0)$. The fundamental problem of causal inference is that for each individual, we only observe a single potential outcome: $Y_i = Y_i(1)\cdot Z_i + Y_i(0)\cdot(1 - Z_i)$. In the context of our applied example, we cannot simultaneously observe what an individual's earnings would have been had they gone to college and what that individual's earnings would have been had they not gone to college, because a person either attends or does not attend. 

An unfortunate consequence of the fundamental problem of causal inference is that the individual causal effects, $\tau_i$, are non-identifiable. While we can not estimate the \textit{individual} causal effects, we can---when necessary structural assumptions are satisfied---estimate \textit{average} causal effects by using observed data to model potential outcomes \parencite{imbens2015causal}. Importantly, there are several distinct average treatment effects. In this work, we focus on the \textit{conditional average treatment effect} (CATE) defined as $\text{CATE} \equiv \mathbb{E}[Y_i(1) - Y_i(0)]$. We highlight the CATE as our average treatment effect of interest because it is the causal estimand most familiar to applied researchers.\footnote{For a review of different average treatment effects, refer to \textcite{RAOS} or  \textcite{hernan2010causal}} Next, we outline the necessary structural assumptions for unbiased estimates of the CATE in observational studies. 

\subsection{Assumptions}

In observational studies, the treatment and control groups differ in their distributions of covariates. We can account for these differences by modeling either the potential outcomes or the assignment mechanism as a function of the observed data \parencite{hill2023machine, rubin1991practical, rosenbaum1983central}. However, this modeling process is only a viable solution when we have measured all confounding variables. In the statistics literature, this unconfoundedness assumption is formalized as the strong ignorability assumption where $Y(1), Y(0) \perp Z\mid\mathbf{X}$, where $\mathbf{X}$ represents all confounding variables \parencite{holland1987causal}. In the present example, we adjust the ignorability notation to $Y(1), Y(0) \perp Z\mid\theta$ where $\theta$ is our latent trait of motivation. An often overlooked aspect of ignorability is that it not only assumes all confounding variables have been measured, \textit{but that they have been measured without error}. This subtlety is central to what we describe as the problem of measurement induced confounding. Here, we assume that motivation is the sole confounding variable. Where we depart from convention, however, is the rejection of the overly optimistic assumption that motivation is measured without error.

When the parametric form of the treatment assignment mechanism and the response surface is unknown, the ability to sufficiently adjust for confounding variables also requires the existence of an empirical counterfactual, meaning that, for every area of the covariate space containing a treated individual, there must be a non-zero probability of containing a non-treated individual (and vice versa). This assumption is known as overlap, common support, or positivity, defined as $0 < P(Z=z\mid\textbf{X}) < 1$, where $\textbf{X}$ represents all confounding variables. Within our context, we denote the overlap assumption as $0 < P(Z=z\mid\theta) < 1$, implying that, for all levels of motivation, there must be at least one individual who attended college and at least one who did not.\footnote{For an in depth exploration of the overlap assumption in observational studies see  \textcite{hill2013assessing}} In all of our forthcoming simulation studies, overlap is known to be satisfied. Moreover, in the context of this work, we also assume a linear parametric relationship for both the treatment mechanism and the response surface, making the overlap assumption unnecessary. 
 
Finally, an implicit assumption of the definition of potential outcomes is that any individual's response to treatment only depends on the treatment assignment of that individual. In practice, this means we must assume that one individual's treatment does not impact another individual's outcome. As such, we must assume that one student attending college does not influence the income of a separate student in our study. Formally, this assumption is described as the stable unit treatment value assumption and abbreviated as SUTVA \parencite{rubin1978bayesian}.\footnote{This assumption applies to both observational and randomized studies} Again, in our subsequent simulations, SUTVA is satisfied.

When the assumptions of strong ignorability, overlap, and SUTVA are satisfied, $\mathbb{E}[Y(0)\mid Z = 1, \,\mathbf{X}] = \mathbb{E}[Y(0)|Z = 0, \,\mathbf{X}]$ and $\mathbb{E}[Y(1)\mid Z = 1,\,\mathbf{X}] = \mathbb{E}[Y(1)\mid Z = 0, \,\mathbf{X}]$. Substituting our single confounder motivation for $\mathbf{X}$ gives $\mathbb{E}[Y(0)\mid Z = 1, \,\theta] = \mathbb{E}[Y(0)\mid Z = 0, \,\theta]$ and $\mathbb{E}[Y(1)\mid Z = 1,\,\theta] = \mathbb{E}[Y(1)\mid Z = 0, \,\theta]$. Under these conditions, estimating the CATE in an observational study becomes a modeling task where the analyst must sufficiently condition on $\theta$ in order to estimate the CATE without bias. As we demonstrate in the following section, the problem of measurement induced confounding occurs when researchers attempt to substitute estimates or proxy measures for $\theta$ without accounting for measurement error.

\section{Measurement Induced Confounding}

 With the motivation as the sole confounding variable, the strong ignorability assumption is that $Y(1), Y(0)\perp Z\mid\theta$. The problem of measurement induced confounding is that $Y(1), Y(0)\not \perp Z\mid\hat\theta^{sum}$ where $\hat\theta^{sum}$ are estimated latent trait values from a sum score measurement model,  $Y(1), Y(0)\not \perp Z\mid\hat\theta^{IRT}$ where $\hat\theta^{IRT}$ are estimated latent trait values from a 2PL IRT measurement model, and $Y(1), Y(0)\not \perp Z\mid X_j$ where $X$ are the item responses themselves. Measurement induced confounding occurs when estimates of latent traits, or proxy measures of latent traits (such as directly adjusting for item responses) are used as substitutes for directly adjusting for the latent trait. The core problem becomes clear when visualized through the language of Directed Acyclic Graphs \parencite[DAG;][]{pearl2009causal, pearl1995bayesian}. The DAG shown in Figure 1 encodes the dependencies between the variables in our hypothetical observational study. Motivation $(\theta)$ is the only confounding variable, responses to the 100-item motivation inventory $(X)$ are a function of an individuals underlying motivation, and estimates of motivation through either sum scoring $(\hat\theta^{sum})$ or a 2PL IRT model $(\hat\theta^{IRT})$ are functions of the observed item responses $(X)$.
 
\begin{figure}
\begin{center}

\begin{tikzpicture}[
    dag_node/.style={circle, draw=black, minimum size=8mm, inner sep=1pt},
    >={Stealth},
    every path/.style={->, thick}
]

    \node[dag_node] (Z) at (0, 2) {Z};
    \node[dag_node] (X) at (2, 4) {X};
    \node[dag_node] (Y) at (4, 2) {Y};
    \node[dag_node] (theta_hat) at (4, 4) {$\hat{\theta}^{IRT}$};
    \node[dag_node] (X_sum) at (4, 5) {$\hat{\theta}^{sum}$};
    \node[dag_node] (theta) at (0, 4){$\theta$};

    \draw (theta) -- (Z);
    \draw (theta) -- (Y);
    \draw (theta) -- (X);
    \draw (X) -- (theta_hat);
    \draw (X) -- (X_sum);
    \draw (Z) -- (Y);

\end{tikzpicture}

\end{center}
\caption{The DAG representing confounding from the latent trait $\theta$ (motivation). Individuals item responses $X$ are a function of the latent trait $\theta$. Sum scores ($\hat{\theta}^{sum}$) and IRT estimates ($\hat{\theta}^{IRT}$) are both functions of item responses $X$.} \label{fig:dag}
\end{figure}

Given the DAG shown above in Figure \ref{fig:dag}, $\theta$ (motivation) must be adjusted for to account for the confounding on the effect of college attendance $(Z)$ on income $(Y)$. Adjusting for items directly$(X)$, sum scores $(\hat\theta^{sum})$, or estimates from an IRT model $(\hat\theta^{IRT})$ does not close the backdoor path through $\theta$, leaving residual confounding. Following the framework introduced by \textcite{textor2016robust}, we can conclude that, given Figure 1, item responses $(X)$, sum scores $(\hat\theta^{sum})$, or IRT estimates $(\hat\theta^{IRT})$ are \textit{not} sufficient adjustment sets. The implication is that adjusting for estimates of latent traits or item responses that are \textit{functions of latent traits} will not eliminate confounding from latent traits $(\theta)$. Measurement induced confounding thus occurs when a researcher mistakenly attempts to use measurements of latent traits to directly account for confounding from latent traits, leaving dependence between the treatment $Z$ and potential outcomes $Y(1)$ and $Y(0)$. 

Due to measurement induced confounding, the only way we can estimate the causal effect of interest is by directly adjusting for the latent trait $\theta$. However, because latent traits are fundamentally not directly observable, we appear to have reached an impasse. As we show in the following section, a Bayesian Joint Estimation approach allows us to directly adjust for the \textit{effect of} $\theta$ without directly observing values of $\theta$ itself.

\section{Bayesian Joint Estimation}

Figure \ref{fig:dag-simple} presents a simplified version the DAG encoding the relationship between our single confounding variable motivation $(\theta)$, item responses from our 100-item motivation inventory $(X)$, treatment variable attending college $(Z)$, and outcome variable income $(Y)$. 

\begin{figure}[ht]
\begin{center}

\begin{tikzpicture}[
    dag_node/.style={circle, draw=black, minimum size=8mm, inner sep=1pt},
    >={Stealth},
    every path/.style={->, thick}
]

    \node[dag_node] (Z) at (0, 2) {Z};
    \node[dag_node] (X) at (2, 4) {X};
    \node[dag_node] (Y) at (4, 2) {Y};
    \node[dag_node] (theta) at (0, 4){$\theta$};

    \draw (theta) -- (Z);
    \draw (theta) -- (Y);
    \draw (theta) -- (X);
    \draw (Z) -- (Y);

\end{tikzpicture}

\end{center}
\caption{This simplified DAG removes $\hat\theta^{sum}$ and $\hat\theta^{IRT}$ because they are not necessary in the Bayesian Joint Estimation approach. Using Bayesian Joint Estimation, we can adjust for the confounding from $\theta$ without directly observing $\theta$. Our approach is directly derived from, and necessarily implied by, this DAG.} \label{fig:dag-simple}
\end{figure}

From this DAG, we write down the full joint distribution,\footnote{For a review of factorization and its application within Bayesian Networks refer to work from Koller and colleagues \parencite*{koller2009probabilistic}.}
\begin{equation}\label{bayesNet}
p(\theta, Z, Y, X) = p(\theta)p(X|\theta)p(Z|\theta)p(Y|\theta, Z).
\end{equation}
Assuming conditional independence across units and local independence across items, we further decompose the joint into
\begin{equation}\label{bayesNetInd}
\begin{split}
&p(\theta)p(X|\theta)p(Z|\theta)p(Y|\theta, Z) \\
&\quad = p(\theta)\prod_{ij}p(X_{ij}|\theta_i)\prod_ip(Z_i|\theta_i)\prod_ip(Y_i|\theta_i, Z_i)
\end{split}
\end{equation}
where $i$ indexes individuals and $j$ indexes items.  

To estimate probabilities from the data, $p(X_{ij}|\theta_i)$, $p(Z_i|\theta_i)$, and $p(Y_i|\theta_i, Z_i)$ are assigned parametric likelihoods and $p(\theta)$ is assigned a prior distribution: 
\begin{align}
&p(X_{ij}|\theta_i) \equiv X_{ij}|a_j, b_j, \theta_i \sim \mathrm{Bern}\Big(
    \operatorname{logit}^{-1}\big( a_j(\theta_{i} - b_j) \big)
    \Big)\label{paramlik1}\\
    &p(Z_i|\theta_i) \equiv Z_i|\beta_0, \beta_\theta, \theta_i \sim \mathrm{Bern}\Big(\operatorname{logit}^{-1}\!\big(
        \beta_0 + \beta_\theta \theta_i
    \big)\Big)\\
    &p(Y_i|\theta_i, Z_i) \equiv Y_i|\gamma_0, \gamma_\theta, \gamma_z, \theta_i,Z_i ,\sigma\sim\mathcal{N}\Big(\gamma_0 + \gamma_\theta \theta_i + \gamma_zZ_i, \sigma^2\Big) \\
    &p(\theta) \sim \mathcal{N}(0, 1)\label{thetaprior}.
\end{align} 
Closer examination reveals that $p(X_{ij}|\theta_i)$ is parameterized as a 2PL IRT model, $p(Z_i|\theta_i)$ is parameterized as a logistic propensity score model where the assignment mechanism is determined by $\theta$, and $p(Y_i|\theta_i, Z_i)$ is parameterized as response surface modeling that adjusts for our single confounding variable, $\theta$. 


Substituting the parametric likelihoods from Equations \ref{paramlik1}--\ref{thetaprior} gives: 
\begin{equation}
\begin{split}
& p(\theta)\times\\&\prod_{ij}\mathrm{Bern}\Big(
    \operatorname{logit}^{-1}( a_j(\theta_{i} - b_j) )
    \Big)\times\\&\prod_i\mathrm{Bern}\Big(\operatorname{logit}^{-1}\!\big(
        \beta_0 + \beta_\theta \theta_i
    \big)\Big) \times \\ &\prod_i\mathcal{N}\Big(\gamma_0 + \gamma_\theta \theta_i + \gamma_zZ_i, \sigma^2\Big)
\end{split}
\end{equation} 
Recall that our inferential goal is to obtain unbiased estimates of the CATE with correctly calibrated measures of uncertainty. Under this parameterization, $\gamma_z$ is our estimate of the CATE. All other parameters are nuisance parameters and, for notational convenance, are collapsed into $\Theta
=
\Big\{
a,\;
b,\;
\theta,\;
\beta_{0},\;
\beta_{\theta},\;
\gamma_0,\;
\gamma_{\theta},\;
\sigma
\Big\}$. Next, we apply Bayes Theorem to obtain the full joint posterior probability of $\gamma_z$ conditional on observed variables $Y$, $Z$, and $X$:  
\begin{equation}
\begin{split}\label{posterior}
p(\gamma_z, \Theta \mid y, z, X)\;\propto\; 
 &p(\gamma_z)p(\Theta)\times \\&\prod_{ij}\mathrm{Bern}\Big(
    \operatorname{logit}^{-1}( a_j(\theta_{i} - b_j) )
    \Big)\times \\&\prod_i\mathrm{Bern}\Big(\operatorname{logit}^{-1}\!\big(
        \beta_0 + \beta_\theta \theta_i
    \big)\Big)\times\\& \prod_i\mathcal{N}\Big(\gamma_0 + \gamma_\theta \theta_i + \gamma_zZ_i, \sigma^2\Big)\\
\end{split}
\end{equation} 
We couple this posterior with weakly informative prior distributions:
\begin{align}
    p(a) &\sim \text{Exp}(1)\\ 
    p(b) &\sim \mathcal{N}(0, 10^2)\\ 
    p(\theta) &\sim \mathcal{N}(0, 1)\\
    p(\sigma)  &\sim \text{Exp}(1) \\
    p(\beta_0), p(\beta_{\theta}), p(\gamma_0), p(\gamma_z), p(\gamma_{\theta})  &\sim \mathcal{N}(0, 5^2).
\end{align}

To sample from the joint posterior $p(\gamma_z, \Theta \mid y, z, X)$, we use the Hamiltonian Monte Carlo with the No U-Turn Sampler (NUTS) implemented in the Stan programming language \parencite{carpenter2017stan, guo2020package}. The Bayesian joint posterior distribution in Equation \ref{posterior} was derived from, and is necessarily implied by, the Bayesian network shown in Figure 2 and was explicitly formulated to estimate the causal effect of a treatment variable in the presence of confounding variables that are latent traits. The key insight is that the DGP implied by Figure 1 and Figure 2 has unmodeled covariance between $X$ and $Z$ due to their common cause, $\theta$. Merely adjusting for summaries of $X$, \textit{regardless of their functional form}, without modeling the assignment mechanism $p(Z_i\mid\theta_i)$ explicitly leads to biased estimates of the CATE, $\gamma_Z$, and thus, undercoverage.

Our approach of explictly modeling the treatment assignment mechanism is not novel and mirrors the Mixed Effects Structural Equations Model described by \textcite{schofield2015correcting} and the Conditional Independence model of \textcite{richardson1993conditional} developed for general modeling contexts. We do provide two key contributions beyond these works, however: First, we explicitly consider the problem of measurement-induced confounding within potential-outcomes based causal inference. Second, we demonstrate the problem (and subsequent solution) through a unique derivation grounded in DAG-based causal models.  

In the following section, we present a simulation study with the aim of providing evidence that measurement induced confounding leads to biased estimates and undercoverage when latent traits are adjusted for with sum scores $(\hat\theta^{sum})$, estimates from a 2PL IRT model$(\hat\theta^{IRT})$, or direct adjustments for item responses $(X)$ and our approach provides unbiased estimates and properly calibrated coverage of the CATE.

\section{Simulation Study}
Our simulation study is designed around our motivating example of an observational study with the aim of estimating the causal effect of college attendance on income. We assume that the researcher knows that motivation is the only confounding variable and that motivation is related to college attendance (the treatment) and income (the outcome) through a simple linear function. To measure motivation, we rely on binary responses from a $J=100$ item motivation inventory. Presented results are based on $M=1000$ replications.\footnote{Full simulation and replication code is available at the accompanying GitHub repository: \url{https://github.com/gperrett/Measurement-Induced-Confounding}} 

\subsection{Data Generating Process}
Equations \ref{eq:dist-theta}--\ref{eq:observed-outcome} show the complete data generating process used to generate item responses (\ref{eq:dist-theta}--\ref{eq:dist-response}), treatment assignments (\ref{eq:assignment-mechanism}--\ref{eq:dist-assignment}), and potential outcomes (\ref{eq:error-treated}--\ref{eq:observed-outcome}). Consistent with our earlier notation, we use $i$ to represent unique individuals. In all simulations, we work with a sample size of $N=1000$ individuals $(i \in 1, 2, 3, \dots, 1000)$. We use $j$ to represent unique items from the 100 item motivation inventory $(j \in 1, 2, 3, \dots, 100)$. Individual motivation is denoted as $\theta_i$ and drawn from a standard normal distribution. Responses to the motivation inventory $(X_{ij})$ are drawn from a 2PL IRT model with difficulty parameter values $(b_j)$ drawn from a standard normal distribution and discrimination parameters $(a_j)$ drawn from a normal distribution with a mean of 1 and standard deviation of .2. The propensity score, also known as the assignment mechanism, $(\pi_i)$ is the probability that an individual will receive the treatment (attend college).

\begin{align}
\theta_i &\sim \mathcal{N}(0, 1^2) \label{eq:dist-theta}\\
b_j &\sim \mathcal{N}(0, 1^2) \label{eq:dist-diff}\\
a_j &\sim \mathcal{N}(1, .2^2) \label{eq:dist-disc}\\
X_{ij} &\sim \mathrm{Bern}\big(\text{logit}^{-1}\big(a_j(\theta_i - b_j)\big)\big)\label{eq:dist-response}\\
\pi_i &= \Phi(\theta_i) \label{eq:assignment-mechanism}\\
Z_i &\sim \mathrm{Bern}(\pi_i)\label{eq:dist-assignment}\\
\epsilon_1 &\sim \mathcal{N}(0, 1^2) \label{eq:error-treated}\\
\epsilon_0 &\sim \mathcal{N}(0, 1^2) \label{eq:error-control}\\
Y_i(1) &= \theta_i +.2 + \epsilon_1 \label{eq:treated}\\
Y_i(0) &=  \theta_i + \epsilon_0 \label{eq:control}\\
Y_i &= Y_i(1)\cdot Z_i + Y_i(0)\cdot(1 - Z_i) \label{eq:observed-outcome}
\end{align} 

An important aspect of the data generating process is that both the assignment mechanism and response surface follow a linear parametric functional form. Recent work in causal inference has focused on flexibly modeling potential outcomes or the assignment mechanism without relying on parametric modeling assumptions \parencite{hill2011bayesian, hill2023machine, dorie2019automated}. In the current context, where the underlying confounding follows a simple linear function, flexible estimators offer no additional benefits and simple linear models should be sufficient to adjust for confounding. Next, we specifically define the nature of these linear models.

\subsection{Estimators}
Our simulation study compares the Bayesian Joint Estimation approach, described and outlined in section 5, against current methods commonly used in social science to adjust for latent traits. While Bayesian Joint Estimation adjusts directly for effects of the unobserved latent trait $\theta$, conventional approaches can only adjust for estimates (or proxy measures) of the underlying latent trait. In our simulation study, we consider adjusting for sum scores $\hat\theta_i^{sum}$, adjusting for an estimate of the latent trait from a 2PL IRT model $(\hat\theta_i^{IRT})$, and directly adjusting for individual item responses themselves $(\mathbf{x}_{i})$.

Equations \ref{eq:surface-sum}--\ref{eq:surface-item} present three estimators that model potential outcomes as functions of the treatment variable and an estimate or proxy for motivation. While each estimator differs in how it measures motivation, they are similar in that they attempt to adjust for confounding by modeling the response surface variable. Moving forward, we refer to this approach as response surface modeling.

\begin{align}
&y_i = \beta_0 + \beta_1 z_i + \beta_2 \hat{\theta}_i^{sum} + \epsilon_i \label{eq:surface-sum}\\
&y_i = \beta_0 + \beta_1 z_i + \beta_2 \hat{\theta}_i^{IRT} + \epsilon_i \label{eq:surface-irt} \\
&y_i = \beta_0 + \beta_1 z_i + \sum_j \beta_jx_{ij} + \epsilon_i \label{eq:surface-item}
\end{align}

An alternative to response surface modeling is modeling the assignment mechanism, the process where individuals self-select into treatment status. Estimating casual effects by modeling the assignment mechanism involves estimating the probability that a given individual receives the treatment, known as the propensity score and commonly denoted as $\pi$. Estimates of this propensity score $(\hat\pi)$ are used to re-weight the data so that data resembles the properties of a randomized experiment. In this work we use inverse probability treatment weighting (IPTW) to re-weight our observational data. If the data is sufficiently re-weighted so that it resembles the balance of randomized studies, a simple difference in means between the treatment and control group will produce an unbiased estimate of the CATE. The IPTW estimator we use for re-weighting and estimation of the CATE is shown in Equation \ref{eq:IPTW}. 

\begin{equation}\label{eq:IPTW}
\frac{1}{n}\sum_i \frac{Y_iZ_i}{\hat\pi_i} -\frac{Y_i(1 - Z_i)}{1-\hat\pi_i}
\end{equation}

The ability for IPTW to produce unbiased estimates of the CATE fully relies on the researchers ability to model the propensity score to sufficiently achieve balance across all confounding variables \parencite{rosenbaum1983central, smith2005does, kang2007demystifying}. As with response surface modeling, researchers using conventional approaches must rely on estimates or proxy measures to model the role of latent traits, such as motivation, in the assignment mechanism. Three approaches for modeling the propensity score are shown in Equations \ref{eq:prop-sum}--\ref{eq:prop-item}. The first uses estimates of motivation from sum scores $(\hat\theta_i^{sum})$, the second estimates motivation with a 2PL IRT model $(\hat\theta_i^{IRT})$, and the third directly adjusts for all item responses $(\mathbf{x}_{i})$. 

\begin{align}
\hat\pi_i &= \frac{1}{1 + \exp\big[-(\beta_0 + \beta_1 \hat{\theta}_i^{sum})\big]}\label{eq:prop-sum}\\
\hat\pi_i &= \frac{1}{1 + \exp\big[-(\beta_0 + \beta_1 \hat{\theta}_i^{IRT})\big]}\label{eq:prop-irt}\\
\hat\pi_i &= \frac{1}{1 + \exp\big[-(\beta_0 + \sum_j \beta_jx_{ij})\big]}\label{eq:prop-item}
\end{align}

A know limitation of IPTW estimators is that they produce biased estimates of standard errors for the CATE \parencite{lunceford2004stratification, austin2016variance}. In our simulations, we correct for this bias by estimating standard errors with a sandwich variance estimator \parencite{williamson2014variance}. Specifically, we use the sandwich estimator in the $\texttt{survey}$ R package \parencite{lumley2020package} as described in Gelman, Hill, and Vehtari \parencite*{RAOS}.

\subsection{Results}

\begin{figure}[ht]
\begin{center}
\includegraphics[width=3in]{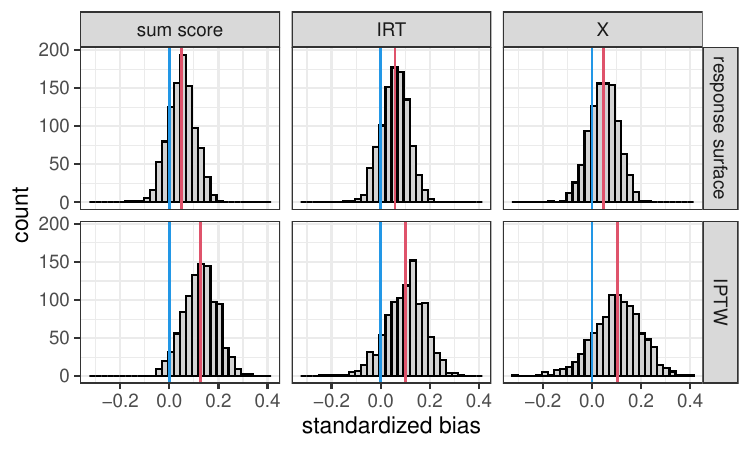}
\end{center}
\caption{Standardized bias for conventional approaches that adjust for estimates from measurement models (sum scores of IRT estimates) or directly adjust for item responses. Top panels display response surface estimators while bottom panels display IPTW estimators. Unbiased estimators converge to 0 (the blue line). Red lines represent average standardized bias for each estimator. Discrepancies between the red and blue lines signal that an estimator is biased. Traditional approaches that attempt to adjust on estimates of latent traits with sum scores or IRT estimates or directly adjusting for item responses all produces biased estimates of the CATE.} \label{fig:bias-standard}
\end{figure}

Standardized bias from response surface modeling and IPTW is presented in Figure \ref{fig:bias-standard}. Across conventional approaches, attempting to adjust for the latent confounding variable (motivation) with sum scores $(\hat\theta^{sum})$, IRT-derived estimates $(\hat\theta^{IRT})$, or item responses directly $(\mathbf{x}_{i})$ all led to biased estimates of the CATE. With response surface modeling, attempting to adjust for confounding with sum scores led to an average bias of .0498 standard deviations, using estimates from a 2PL IRT model did not lead to improvement and resulted in an average bias of .0584 standard deviations, and directly adjusting for observed item responses produced an average bias of 0.0475 standard deviations. Standardized bias was even larger under IPTW models, estimating the propensity score from sum scores lead to an average standardized bias of .126, estimating the propensity score with estimates from 2PL IRT model led to an average standardized bias of .102, and estimating the propensity score with observed item responses led to an average standardized bias of .103.

\begin{figure}
\begin{center}
\includegraphics[width=3in]{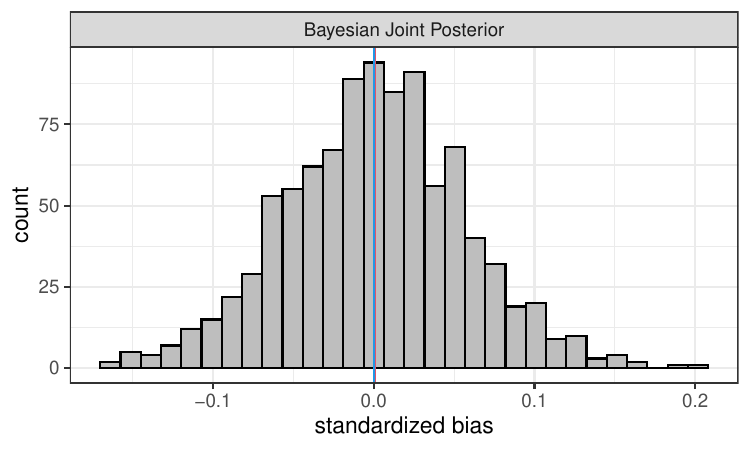}
\end{center}
\caption{Standardized bias from the Bayesian Joint Estimation approach. Under this approach estimates converge to 0 standardized bias and the red and blue lines overlap. Unlike conventional approaches, Bayesian Joint Estimation can recover unbiased estimates of the CATE in the presence of confounding from a latent trait.} \label{fig:bias-bje}
\end{figure}

Figure \ref{fig:bias-bje} presents results from our proposed Bayesian Joint estimation approach outlined above. Conventional estimation approaches that attempt to directly adjust for measurements of the latent confounder led to biased estimates of the Average Treatment Effect, but when the latent trait, assignment mechanism, and response surface are modeled jointly we obtained unbiased estimates of the Average Treatment Effect.

\begin{figure}
\begin{center}
\includegraphics[width=3in]{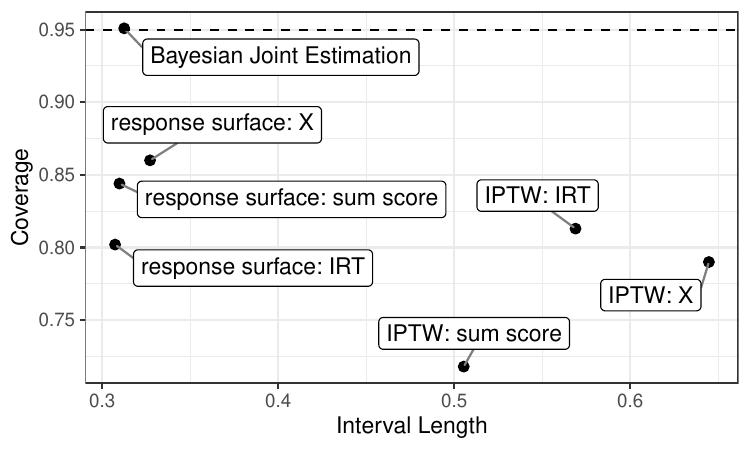}
\end{center}
\caption{Coverage rates ($y$-axis) and average 95\% uncertainty interval length ($x$-axis) for all estimation methods. The Bayesian Joint Estimation approach achieves nominal coverage while conventional approaches all undercover. The average interval length for the Bayesian Joint Estimation approach is equal to or less than all other approaches demonstrating that nominal coverage does not come at the expense of efficiency.} \label{fig:coverage}
\end{figure}

Well calibrated uncertainty intervals are equally, if not more important, than estimates that are unbiased. Coverage of 95\% uncertainty intervals for our Bayesian Joint estimation and conventional approaches are presented in Figure \ref{fig:coverage}. The Bayesian Joint estimator is perfectly calibrated, with 95\% intervals covering the true average treatment effect 95\% of the time. In contrast, all other approaches systematically undercovered. Critically, the coverage properties of our Bayesian Joint estimation approach did not come at the expense of greater uncertainty. The length of the 95\% uncertainty intervals was comparable or shorter than alternative approaches signaling that the desirable coverage properties are from better estimation rather than from inflating uncertainty. 

\section{Evidence from a Constructed Observational Study}
A criticism of simulation studies is that they are unrepresentative of actual data analysis in applied settings \parencite{hernan2019comment}. To supplement our simulation-based results, we turn to a constructed observational study developed by \textcite{keller2025new}. Here, participants were enrolled in a randomized study on the causal effect of a mathematics training session on a subsequent math test (a vocabulary training session served as the control arm). Prior to assignment, participants were asked if they would prefer to enroll in a mathematics training session or a vocabulary training session. To construct the parallel observational study, Keller and colleagues dropped participants whose randomized treatment assignment did not align with their preferences. After this alteration, participants in the observational dataset have self selected into their preferred treatment arms. Results from the randomized study thus provide an unbiased experimental benchmark for the true CATE, while the constructed observational data provide a testing ground for observational methods where the goal is to recover the effect from the randomized study when using the observational data. 

In the constructed observational study, we consider gender, race, marital status, age, income, educational attainment, parents' educational attainment, exposure to a calculus course, whether a participant endorses ``I like Math,'' and baseline mathematical ability as confounding variables. Baseline mathematical ability is treated as a latent variable measured by a 12 item pretest exam. While most observational studies fail to report pre-intervention measures of potentially confounding latent variables at anything finer than an aggregate level, this analysis is enabled by \textcite{keller2025new} providing all individual item-level responses. 

\begin{figure}[ht]
\begin{center}
\includegraphics[width=3in]{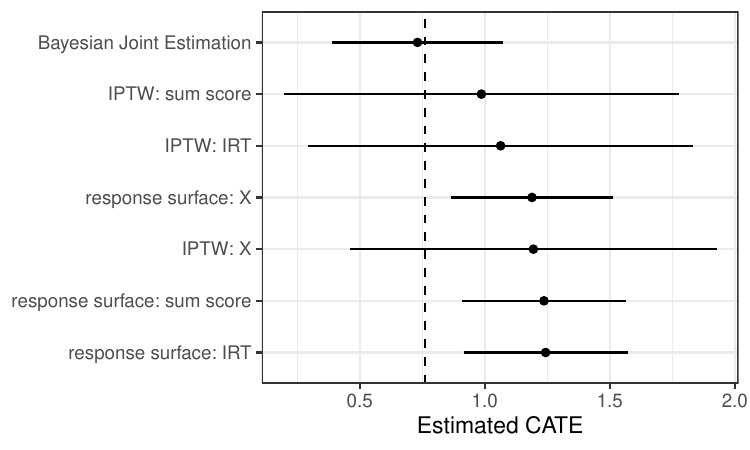}
\end{center}
\caption{All estimation methods ($y$-axis) and the estimated CATE ($x$-axis) for the constructed observational study. The dashed line represents the unbiased benchmark from the randomized study. Estimation methods are ordered by bias (relative to the experimental benchmark). The Bayesian Joint Estimation approach recovers the result of the randomized study. All other approaches are more biased and approaches that rely on modeling the response surface all have 95\% uncertainty intervals that fail to cover the experimental benchmark.} \label{fig:obs-result}
\end{figure}

Figure \ref{fig:obs-result} presents results from our analysis. We compare estimates of the CATE from the Bayesian Joint Estimation approach against the IPTW and response surface approaches considered in the previous section. These results demonstrate a real applied research setting in which adjusting for a latent confounding variable (baseline mathematics ability) using sum scores, IRT-derived scores, or individual item responses is used to control for confounding.

The experimental benchmark yields a CATE of .7603. Our Bayesian Joint Estimation approach finds a CATE of .73, almost identical to the experimental benchmark. In contrast, conventional response surface modeling approaches yielded estimates of 1.24 when mathematical ability is adjusted for using sum scores, 1.24 when mathematical ability is adjusted for using IRT scores, and 1.19 when raw item responses are used to adjust for mathematical ability. When using IPTW, estimates were .99 when adjusting with sum scores, 1.06 when adjusting with IRT scores, and 1.19 when adjusting with raw item responses. 

To better understand the magnitude of these differences, we reframe these results in terms of standard deviations of the outcome variable. Here, conventional approaches ranged from a low of .06 standard deviations away from the experimental benchmark (IPTW: sum score) to a high of .132 standard deviations away from the experimental benchmark (response surface: IRT). In sharp contrast, the estimate from Bayesian Joint Estimation is only .007 standard deviations away from the experimental benchmark. Notably, these results are nearly identical the standardized bias produced by conventional approaches in our simulation study (see Figures \ref{fig:bias-standard} and \ref{fig:bias-bje}). 

The 95\% uncertainty intervals from all response surface models failed to cover the experimental benchmark. While all 95\% uncertainty intervals from IPTW estimators covered the experimental benchmark, this finding was driven by the high degree of uncertainty (wider intervals). In practice, researchers aim for estimates that are both unbiased and efficient (low uncertainty). Here, IPTW estimates are both far away from the benchmark (indicative of bias) and inefficient (wide intervals). Bayesian Joint Estimation is the only method presented here that produces an efficient estimate that is nearly identical to the experimental benchmark.

\section{Discussion}

Fields such as education and psychology often seek to do causal inference that requires adjusting for latent confounding variables. The most pressing implication of our findings is that observational studies that adjust for latent traits (regardless of whether they are using IRT estimates, sum scores, or item responses) when estimating average treatment effects likely yield biased estimates with incorrectly calibrated uncertainty intervals. Our results show that by simultaneously estimating a measurement model for latent traits, the treatment assignment model, and the response surface, we can obtain unbiased estimates of the CATE in the presence of a latent confounding variable. Importantly, our joint estimation approach does not come at the expense of estimation efficiency.

In practice, researchers often adjust for numerous confounding variables rather than a single confounder. While our hypothetical example considered the case of a single latent confounding variable (motivation), the Bayesian Joint Estimation approach can trivially scale to adjust for multiple latent confounding variables by adding an additional measurement model for each latent confounding variable. Given $K$ latent confounders, the joint model is fit with $K$ measurement models, but is always fit with one treatment assignment model and one response model. An example of a setting with multiple latent confounding variables is provided in the supplemental GitHub repository.\footnote{See: \url{https://github.com/gperrett/Measurement-Induced-Confounding}}  Similarly, the Bayesian Joint Estimation approach proposed in this work also readily accommodates adjustment for variables that are \textit{not} latent. Our analysis of the constructed observational study requires also controlling for numerous directly observable confounders. For this adjustment, observed confounding variables should be included in the treatment assignment and response models. Because these variables are directly observed, a separate measurement model is not necessary. 

A substantial limitation of the proposed Bayesian Joint Estimation is the reliance on strong parametric assumptions. Even when all confounding variables are adjusted for, failing to account for existing nonlinearities and interaction effects among them leads to biased estimates of average treatment effects and poorly calibrated coverage \parencite{hill2011bayesian, hill2023machine}. In settings without latent confounding variables, nonlinearities and interactions are easily accounted for by estimating both the treatment assignment and response surface with flexible machine learning models that minimize parametric modeling assumptions \parencite{dorie2019automated}. Critically, existing flexible estimation methods do not yet solve the problem of measurement induced confounding. As with the parametric cases presented in this paper, under the currently available flexible estimators (e.g., Bayesian Additive Regression Trees \parencite{chipman2010bart, hill2011bayesian}, Causal Forests \parencite{athey2019estimating}, and SuperLearner \parencite{polley2011super}), the only latent variable specifications available to researchers are to either adjust for estimates of latent traits or item responses directly. As we have shown in the linear parametric setting, these approaches are insufficient to adjust for the confounding from latent traits. We know that flexible models alone will not solve the problem of measurement induced confounding, because their key advantage is that they do not require knowledge of the correct parametric model \textit{a priori} and do nothing to address the backdoor path through the assignment mechanism. In this work, all conventional estimation approaches fail even under the correct parametric specification. This result implies that existing nonparametric methods for causal inference cannot address confounding from latent variables. 

The ability of researchers to adjust for latent confounders without making strong parametric assumptions is critical for unbiased estimates of average treatment effects in observational studies where confounding by latent traits is present. In our analysis of the constructed observational study by \textcite{keller2025new}, we were able to recover the experimental benchmark while making strong linear parametric assumptions, however, evidence from the pioneering LaLonde constructed observational study shows that relying on strong parametric assumptions can lead to biased estimates in observational settings \parencite{lalonde1986evaluating, dehejia2002propensity, imbens2024lalonde}. Extending the Bayesian Joint Estimation approach demonstrated in this work to nonparametric models introduces substantial complexity, but is an important area of future research. In the linear setting demonstrated in this work, joint estimation of the measurement model, treatment assignment model, and response surface requires computing partial derivatives of the log posterior with respect to the parameters \parencite{betancourt2017conceptual}. Bayesian Additive Regression Trees \parencite[BART;][]{chipman2010bart} are a natural choice for a Bayesian nonparametric estimator for causal inference problems, because BART simultaneously provides flexible estimation and uncertainty quantification. The primary challenge of extending Bayesian joint estimation to a BART-type model is that BART estimates are obtained as a sum of step functions, which are inherently non-continuous and non-differentiable. This problem is not BART specific. Other popular nonparametric estimators for causal inference, such as causal forests \parencite{athey2019estimating}, are also based on non-differentiable step functions. One potential path forward is to expand upon the \texttt{stan4bart} implementation \parencite{dorie2022stan}, which combines the Stan and BART models within a single Gibbs sampler; however, the current implementation of \texttt{stan4bart} cannot accommodate an IRT-derived measurement model specification.\footnote{Note that while Stan can, at the time of this writing, easily accommodate bespoke IRT models, \texttt{stan4bart} cannot.}

A separate parametric concern in the present work is our reliance on a 2PL IRT model to estimate latent traits. We selected a 2PL model because it is a flexible IRT specification that performs well even when the underlying data generating process for item responses is not a 2PL model \parencite{domingue2024intermodel}. In cases of unipolar, asymmetric, or other bespoke response models, users could supply a different measurement model or look to a non-parametric alternative \parencite{huang2025theory, feuerstahler2026defining, sijtsma1998methodology}.

Given that many observational studies adjust for sum scores on various latent traits, an uncomfortable implication of the results of this work is these studies likely report biased estimates of average treatment effects with incorrectly calibrated uncertainty intervals. Fortunately, the problem of measurement induced confounding is inherently limited to observational studies. We have presented a solution to the problem of measurement induced confounding when linear parametric models are sufficient approximations of the data generating process, however, an alternative, and more optimal, solution is to run randomized studies. Under random treatment assignment, the distribution of all pre-treatment variables, including latent variables, are identical in expectation between the treatment and control group \parencite{RAOS}. 

An implication of this property is that under random assignment, the ignorability assumption is $Y(1), Y(0) \perp Z$, rather than the stronger form of $Y(1), Y(0) \perp \textbf{X}$ necessary in observational studies. When randomization is pristine, there is no loss to follow up, and SUTVA is satisfied, there are no confounding variables because all pre-treatment variables are independent from treatment assignment and unbiased estimates of average treatment effects can be obtained without additional statistical adjustment. Randomized studies resolve the problem of measurement induced confounding because randomization eliminates all confounding variables by design \footnote{This is assuming the structural assumptions that $Y(1), Y(0)\perp Z$ and SUTVA are satisfied}.

While randomization address the problem of measurement induced confounding by design, running randomized studies is not always possible. We choose our hypothetical example, the influence of college attendance on income, because it is a clear example of a setting where randomization is impermissible. Even when randomization is ethically defensible, randomized studies still poss logistical constraints and in some contexts observational studies are the only logistically feasible option. For researcher who must rely on observational studies, this work cautions against relying on traditional measurement models to adjust for confounding from latent traits. 


\printbibliography

\end{document}